\documentclass[sigconf]{acmart}

\renewcommand\footnotetextcopyrightpermission[1]{} 
\usepackage{multirow}
\usepackage{dblfloatfix}
\usepackage{tikz}
\usepackage{pgfplots}
\usetikzlibrary{backgrounds, positioning, fit}
\usetikzlibrary{shapes.geometric}
\usepackage{amsmath}
\usetikzlibrary{patterns}
\usetikzlibrary{pgfplots.groupplots}
\usepackage{subfigure}
\usepackage{float}
\usepackage{wrapfig}

\newif\ifcomment
\commenttrue
\ifcomment
\newcommand{\virat}[1]{{\color{red} \textbf{Virat: #1}}}
\else
\newcommand{\virat}[1]{}
\fi

\begin{document}
\pagestyle{plain}

\title{Quantifying Privacy Leakage in Graph Embedding}

\author{Vasisht Duddu}
\affiliation{%
  \institution{Univ Lyon, INSA Lyon, Inria, CITI}
}
\authornote{Work partly done at author's current affiliation with University of Waterloo, Canada}
\email{vasisht.duddu@uwaterloo.ca}

\author{Antoine Boutet}
\affiliation{%
	\institution{Univ Lyon, INSA Lyon, Inria, CITI}
}
\email{antoine.boutet@insa-lyon.fr}

\author{Virat Shejwalkar}
\affiliation{%
	\institution{University of Massachusetts Amherst}
}
\email{vshejwalkar@cs.umass.edu}

\begin{abstract}

Graph embeddings have been proposed to map graph data to low dimensional space for downstream processing (e.g., node classification or link prediction). With the increasing collection of personal data, graph embeddings can be trained on private and sensitive data. For the first time, we quantify the privacy leakage in graph embeddings through three inference attacks targeting Graph Neural Networks. Our \emph{membership inference attack} aims to infer whether a graph node corresponding to an individual user's data was a member of the model's private training data or not. We consider a \emph{blackbox} setting where the adversary exploits the output prediction scores and a \emph{whitebox} setting where the adversary has also access to the released node embeddings. Our attack provides accuracy up to 28\% (blackbox) and 36\% (whitebox) beyond the random guess by exploiting the distinguishable footprint between train and test data records left by the graph embedding. In our \emph{graph reconstruction} attack, the adversary aims to reconstruct the target graph given the corresponding graph embeddings. Here, the adversary can reconstruct the graph with more than 80\% of accuracy and infer the link between two nodes with $\sim$30\% more accuracy than the random guess. Finally, we propose an \emph{attribute inference attack} where the adversary aims to infer the sensitive node attributes corresponding to an individual user. We show that the strong correlation between the graph embeddings and node attributes allows the adversary to infer sensitive information (e.g., gender or location).

\end{abstract}


\maketitle

\section{Introduction}\label{introduction}

Large scale real-world systems are typically modeled in the form of graphs: mobility, network infrastructures, online social networks, world wide web, citation networks and biomedical datasets, which represent the entities as nodes and their relationship with edges~\cite{zhou2018graph}.
Traditional Deep Neural Networks fail to capture the nuances of structured data but a specific class of algorithms, namely, \emph{Graph Neural Networks} (GNNs) have shown state of the art performance on such complex graph data for node classification, link prediction, etc.
An important pre-processing step for using graph data with machine learning is embedding the high dimensional graph data to a low dimensional representation for easy processing by machine learning algorithms.
In many applications, such embeddings are released for further processing to save storage costs, however, without considering the privacy implications.
Such large graph datasets raise the question of privacy of the underlying training data, specifically if the algorithms are trained with private and potentially sensitive data.

In the context of Machine Learning (ML), a privacy violation occurs when an adversary infers something about a \emph{particular} user's data record in the training dataset which cannot be inferred from other models trained on similar data distribution~\cite{membershipinf,whitebox}.
This information leakage is quantified using the success of various inference attacks. 
In attribute inference attacks, the attacker infers sensitive features of an individual's data record used in the model's training.
A stronger case of attribute inference is where the attacker reconstructs a portion of the private training data, i.e., data reconstruction attack.
In membership inference attacks, the adversary infers the membership of a particular individual's record in the private training data.
Prior literature in privacy attacks focuses on the models trained on non-graph data including text, images, and speech to study the vulnerability to membership inference~\cite{ndss19salem,membershipinf}, attribute inference~\cite{attributeinf,attributeinf2,dysan}, property inference~\cite{propertyinf}, model inversion~\cite{modelinversion} attacks as well as model parameter and hyperparameter stealing attacks~\cite{timing,stealml,8418595}.
While well studied in traditional ML, the privacy risk in graph-based ML models under adversarial setting has not been fully explored and quantified. 

Consider a graph that captures the outbreak of a disease where the nodes represent the users, node features indicate medical symptoms, and the edges indicate the disease transmission.
Typically, in such datasets, a GNN provides state of the art performance for predicting disease for an arbitrary user in the graph (via node classification) and determining the future outbreak (via link prediction).
However, when such embedding models do not account for privacy, an adversary can infer the health status of a particular user by identifying whether the user was part of the training data or not.
Further, the adversary can potentially reconstruct the sensitive input graph from the low dimensional embeddings. 
Finally, graph embeddings capture important semantics from the input graph while maintaining the contextual information in the form of preferential connection which can be exploited to infer sensitive attributes about an individual.
These three privacy attacks, namely, membership inference, graph reconstruction and attribute inference, are examples of direct privacy violations of the users which can further be used without user consent. 
Further, companies spend enormous resources to annotate the training dataset to achieve state of the art performances and such attacks which infer training data (graph) also violate the Intellectual Property.


In this work, we propose the first comprehensive privacy analysis of graph embedding algorithms under different threat models and adversary assumptions.
We focus on the privacy analysis of publicly released graph embeddings that are trained with private graph data and used for different downstream tasks. For the privacy analysis, we use various attacks that violate users' privacy: membership inference, graph reconstruction, and attribute inference.
First, we evaluate the privacy leakage using membership inference attacks where the adversary has access to (a) output predictions of GNNs (blackbox setting) and (b) graph embeddings (whitebox setting).
The blackbox setting considers the specific case of downstream node classification tasks for convolution kernel-based graph embedding with neural networks.
We propose two blackbox membership inference attacks: with auxiliary knowledge on the data distribution, called \emph{shadow model attack}, and without auxiliary knowledge, called \emph{confidence score attack}.
Here, we show that the proposed attacks have an inference accuracy of 78\%, 63\%, and 60\% for confidence score attack and 62\%, 60\%, and 55\% for shadow model attack, respectively for three different datasets, i.e., Cora, Citesser and Pubmed dataset.
For the whitebox setting, we propose an unsupervised attack for the more generic case of using just the graph embeddings to differentiate whether a given node was part of the training graph or not.
We show that an adversary in this setting can predict the training data with a high accuracy (70\% on average on the three datasets).

Second, we propose a novel graph reconstruction attack where the adversary, given access to the node embeddings of a subgraph, trains an encoder-decoder model to reconstruct the target graph from its publicly released embeddings.
This attack has serious privacy implications since the adversary reconstructs the input graph which can be potentially sensitive.
This attack has high precision: 0.722 for Cora, 0.778 for Citeseer and 0.95 for Pubmed dataset.
Moreover, on increasing the adversary's prior knowledge, the attack performance increases significantly.
An important privacy implication is the \emph{link inference}, where the adversary predicts the link between any two target nodes in the graph.
Through this attack, an adversary infers the link between the target nodes with 80\% of accuracy on average, compared to the 50\% baseline random guess accuracy.

Finally, we propose an attribute inference attack where the adversary tries to infer sensitive attributes of the user node in the graph using the released graph embeddings.
We consider two state of the art unsupervised random walk based embeddings, Node2Vec and DeepWalk, on two real-world social networking datasets: Facebook and LastFM, where the adversary aims to infer the user gender and location, respectively.
Given access to the embeddings of a subgraph and corresponding sensitive attributes, we model attribute inference as a supervised learning problem.
The adversary trains a supervised attack model to predict sensitive, hidden attributes for target users given the publicly released target embeddings.
Here, F1 score of the attack (which captures the balance between precision and recall) for LastFM (Facebook) dataset was as high as 0.65 (0.59) for DeepWalk and 0.83 (0.61) for Node2Vec.
Our work highlights serious data privacy risks in graph data processing algorithms and calls for further research to design privacy-preserving embedding algorithms for graph data.
Our results can be easily reproduced using the our code\footnote{\url{https://github.com/vasishtduddu/GraphLeaks}}.

The paper organization is as follows.
Section~\ref{background} gives background on graph embedding algorithms and GNNs are introduced, Section~\ref{attack} details our threat models. 
Section~\ref{setup} describes experimental setup.
Section~\ref{evaluation} gives results of our evaluations and Section~\ref{related} discusses related work.
We conclude our work in Section~\ref{conclusions}.

\section{Background}\label{background}

A large number of real-world applications require processing graph data which contains rich relational information between different entities (e.g., online social media, disease outbreaks, recommendation engines, knowledge graphs and navigation systems)~\cite{zhou2018graph}.
Deep Learning and more precisely Convolutional Neural Networks have shown tremendous performance over non-graph data, e.g., image classification, by capturing the spatial relation between pixels of an image and extracting features over multiple layers.
However, these machine learning algorithms are ineffective to learn on graph data~\cite{zhou2018graph}.
Indeed, the models have to capture the connections in the graph data while ensuring invariance of graph data representations, even without fixed ordering between the nodes. In other words, the adjacency matrix representing the connections between nodes varies but still results in the same graph. 
To overcome this limitation, the graph data is passed through embedding algorithms that map the large graphs to lower dimensions which are then used for downstream processing with GNNs.
Graph embedding algorithms enable models operating on low dimensional euclidean datasets (e.g., images) to graph data by mapping them into a low dimensional embedding.
We represent a graph as $G=(V,E)$ where $V$ represents the vertex set consisting of nodes \{$v_1,...,v_n$\} and $E$ represents the edges between the nodes. $E$ is represented as a symmetric, sparse adjacency matrix $A$ $\in$ $R_{nxn}$ where $a_{ij}$ denotes the edge weight between nodes $i$ and $j$ with $a_{ij}= 0$ for missing edges.

\subsection{Graph Embedding Algorithms}

To mitigate the space and computation overhead of downstream graph processing, graph embedding algorithms provide an efficient approach to represent the graph data in a low dimension embedding~\cite{tutorial}.
Specifically, an embedding algorithm $\Psi: V \rightarrow \mathcal{R}^d$, where nodes $V$ $\in$ $G$, takes a node $v\in V$ as its input and outputs a  $d$-dimensional vector that captures the properties of the original graph, e.g., distance of $v$ from other nodes in $G$.
Different graph embedding algorithms to embed both the entire graphs as well as the nodes are well studied~\cite{survey}.
Random walk based node embedding algorithms traverse the graph to sample random walks which are then passed as sentences to SkipGram algorithm to obtain the corresponding node features~\cite{node2vec,deepwalk}.
In deep learning-based graph embeddings, both the features along with adjacency matrix can be used to generate low dimensional embeddings for each of the nodes.
For generating these embeddings, the parameters of the embedding function are updated to improve the representation of the graph nodes while maintaining the original properties.
These are typically modeled using GNNs and Graph Convolutional Networks.
In this work, we focus on node embeddings and refer to them as graph embeddings.
We consider random walk based embeddings for attribute inference and deep learning-based embeddings for node inference and graph reconstruction attacks.

\subsection{Graph Neural Networks}

The initial layers of a GNN are used to generate embedding for the input graphs which can be extended for node classification and link prediction tasks by attaching a classifier network as GNNs.
The pre-processed graph in the form of embeddings along with the node features are represented as matrices for computation.
The training of GNNs relies on a message-passing algorithm which is the weighted aggregation of features of neighboring nodes $\mathcal{N}(v)$ to compute the feature of a particular node $v$.
Given the features $x$ of a single node $v$, the GNN produces an output label $f(x;W)$ which captures the probability of the input node with features $x$ belonging to a particular class.
The loss over the resultant classification for the node $v$ is then backpropagated to update the model weights for aggregation.
Consider a $N\times D_F$ feature matrix $X$, where $N$ is the number of nodes and $D_F$ is the number of node features. Consider an adjacency matrix $A$ which captures the representation of graph structure in matrix form.
The output of a layer with $F$ features takes the feature matrix along with the adjacency matrix as input to produce a $N\times F$ matrix as an output.
The computation is given by $H^{(l+1)} = f_{agg}(H^{(l)}, A)$ with $H(0)=X$ and $H(L)=Z$, $L$ being the number of layers and $H$ is the intermediate activation.
Based on the different aggregation function $f_{agg}()$, we obtain different GNN algorithms such as Graph Convolutional Network (GCN)~\cite{Kipf2016tc}, GraphSAGE \cite{NIPS20176703}, Graph Attention Networks (GAT)~\cite{velickovic2018graph} and Topology Adaptive GCN (TAGCN)~\cite{du2018topology}.


\section{Threat Models}
\label{attack}

In this section, we describe the threat models of various attacks we study, the methodology of the attacks, and adversary assumptions.

\subsection{Membership Inference Attacks}
\begin{figure}[!htb]
\centering
\includegraphics[width=0.85\linewidth]{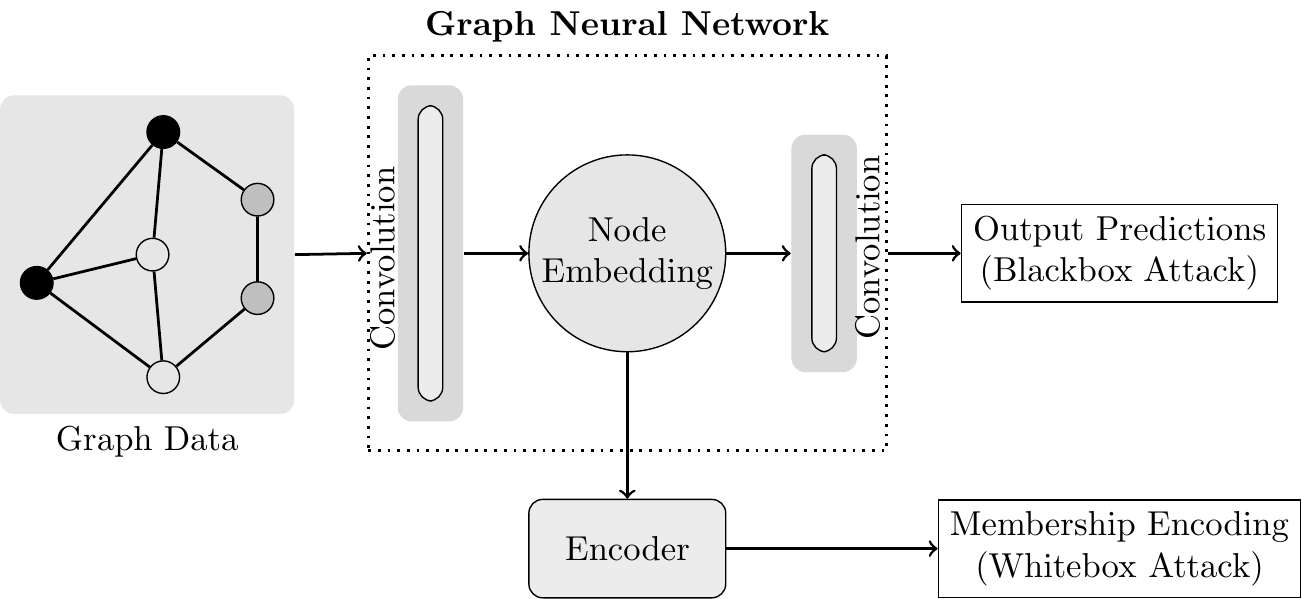}
\caption{Blackbox and whitebox inference attacks to distinguish members and non-members of $G_{train}$.}
\label{mia}
\end{figure}

Membership inference attacks allow personal information leakage in GNNs.
Specifically, the goal of the adversary is to identify whether a user node $v$ is part of the graph $G_{train}$ used for training the target model.
This is a binary classification problem where the adversary learns the threshold to predict the membership of a user node.
Depending on the adversary's knowledge about $f()$, we consider two settings: blackbox (with and without auxiliary knowledge) and whitebox. 
As shown Figure~\ref{mia}, to distinguish between members and non-members of training graph $G_{train}$, the blackbox attacks exploit the statistical difference in output predictions while the whitebox attack exploits the intermediate low dimensional embedding.

\subsubsection{Blackbox: Inference using Output Predictions}

In this setting, we consider the target model as a trained GNN for node classification tasks.
The adversary aims to infer whether a user's node in the graph was used in training the target model $f()$.
In a blackbox setting, the adversary has only access to the model outputs $f(x;W)$ for a given input $x$.
The parameters of the trained model $W$ as well as the intermediate computation are inaccessible to the adversary.
This is a practical setting, typically seen in the case of Machine Learning as a Service, where a trained model is deployed in the cloud and the adversary queries the model through an API and receives corresponding predictions.

\begin{figure}[!htb]
    \centering
    \begin{minipage}[b]{1\linewidth}
    \centering
    \subfigure[Citeseer]{
    \label{fig:nonmem_soft_label}
    \includegraphics[width=0.48\linewidth]{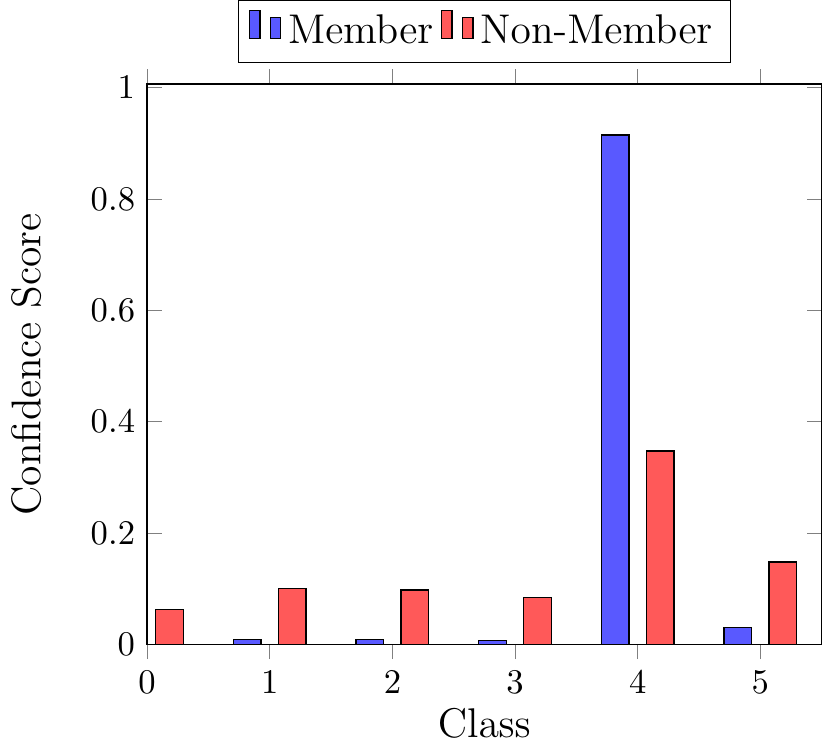}
    \raisebox{2mm}{\includegraphics[width=0.5\linewidth]{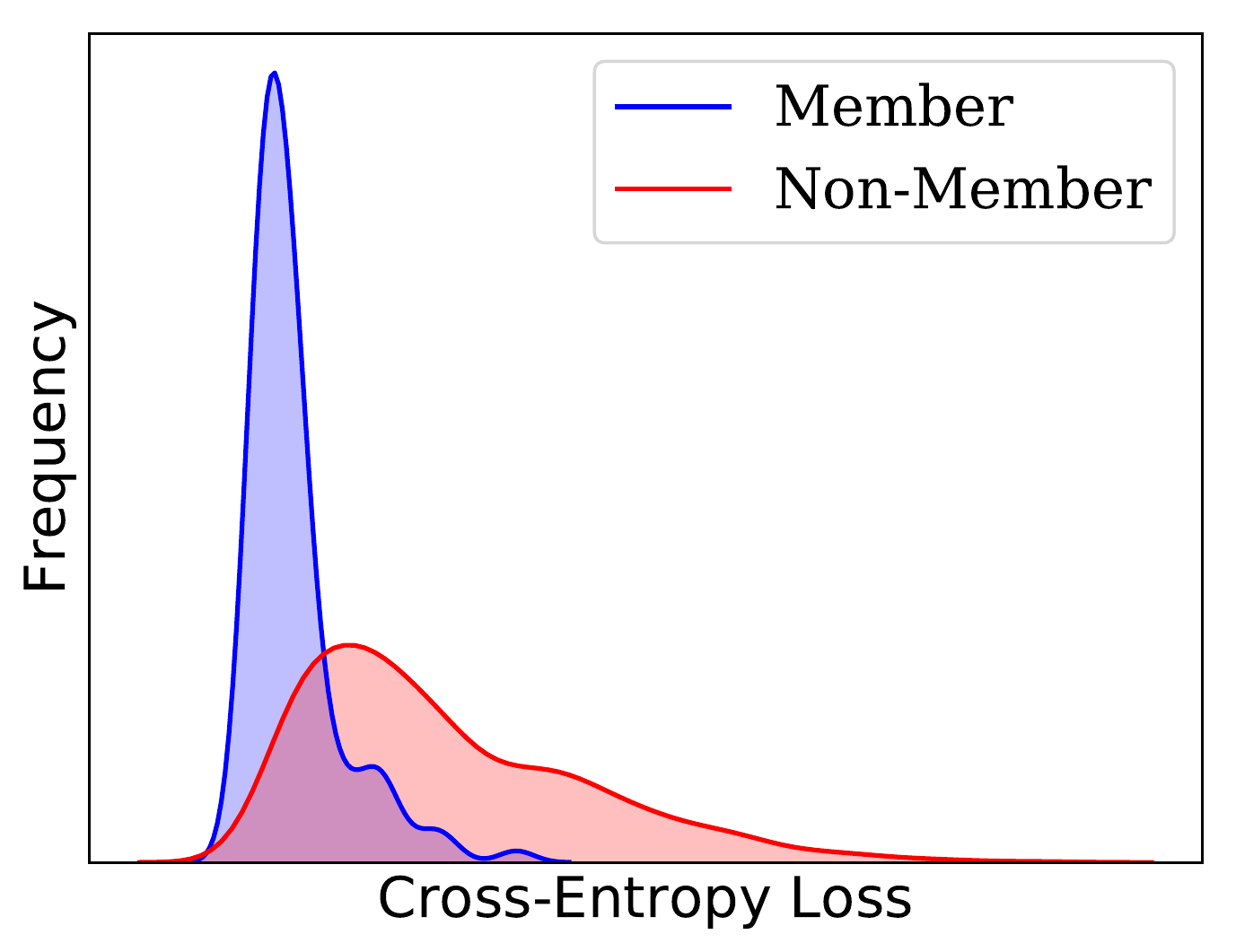}}
    }

    \subfigure[Cora]{
   	\label{fig:mem_soft_label}
    \includegraphics[width=0.48\linewidth]{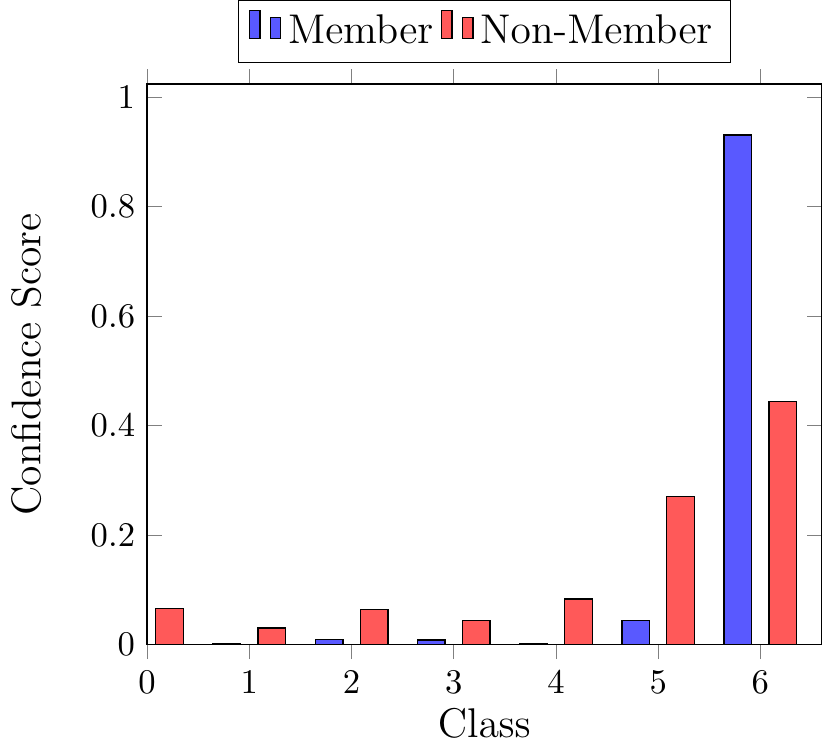}
    \raisebox{2mm}{\includegraphics[width=0.5\linewidth]{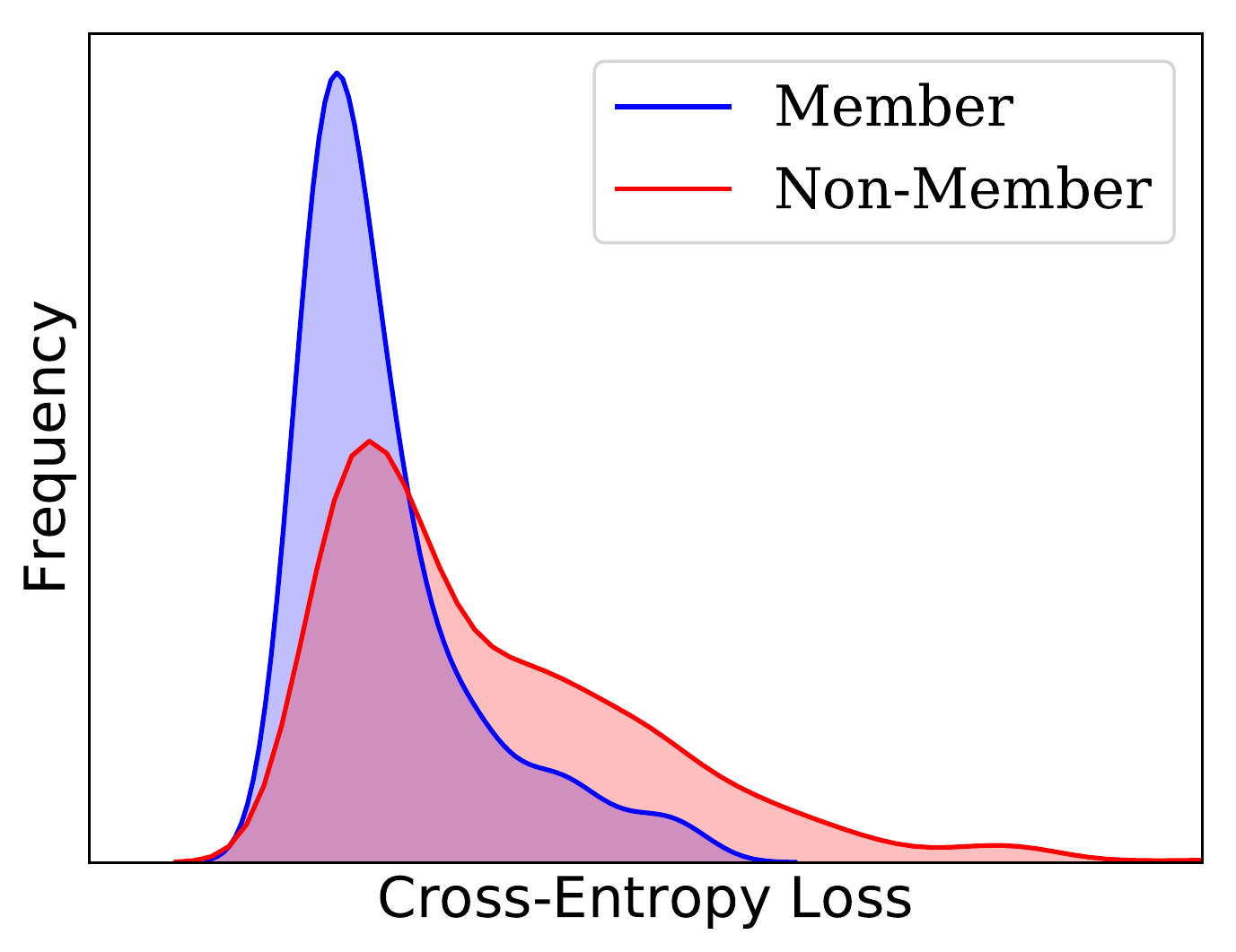}}
    }

    \subfigure[Pubmed]{
    \label{fig:mem_soft_label}
    \includegraphics[width=0.48\linewidth]{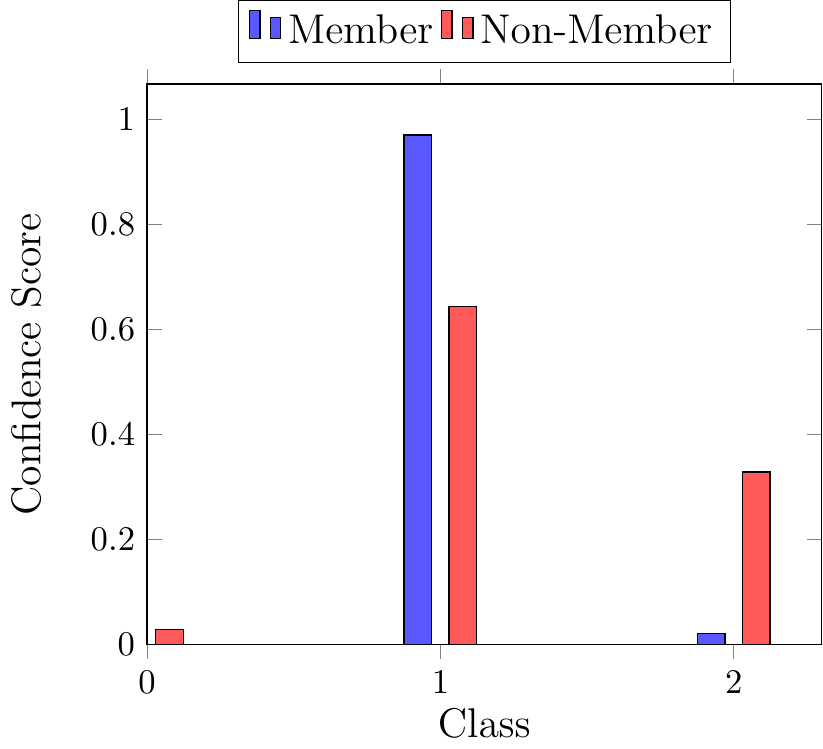}
    \raisebox{2mm}{\includegraphics[width=0.5\linewidth]{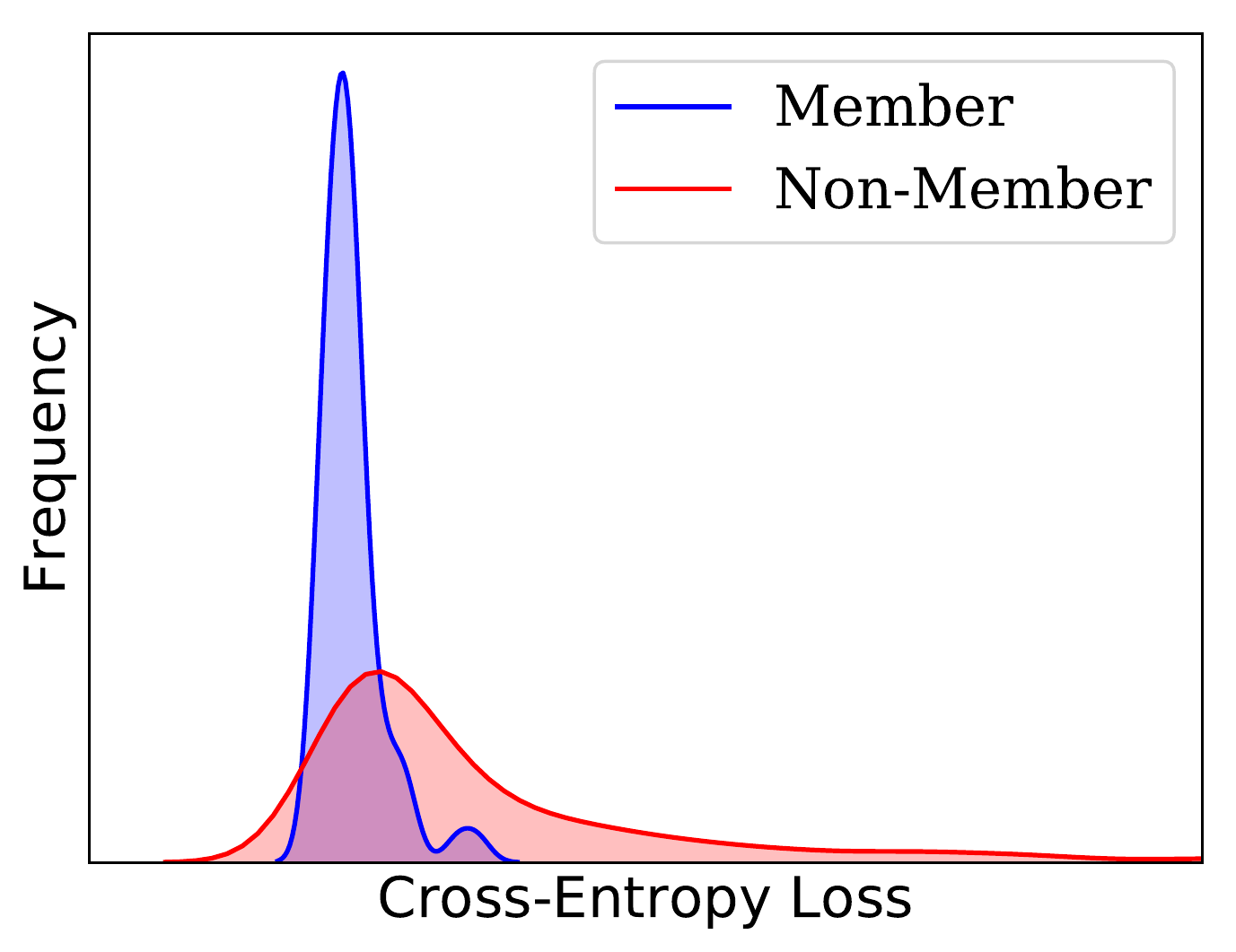}}
    }
    \end{minipage}
    \caption{Model predictions are more confident for nodes in $G_{train}$ compared to test graph (left). The extent of overfitting can be detected by a non-overlapping region between the output prediction distributions across all data points (right).}
    \label{fig:NIAcause}
\end{figure}

Blackbox adversary exploits the statistical difference between the confidence in prediction on training and testing data~\cite{membershipinf}. 
Figure~\ref{fig:NIAcause} (left) illustrates this difference where the prediction confidence for one class is much higher for training data points.
Predicting with higher confidence in seen $G_{train}$ nodes compared to unseen test nodes is referred to as overfitting.
This difference in the output prediction confidence directly results from a distinguishable output distribution between the train and test data indicated by a non-overlapping region between distributions (Figure~\ref{fig:NIAcause}, right).

We consider two attacks within the blackbox setting: (a) shadow model attack and (b) confidence score attack.

\begin{figure}[!htb]
\centering
    \includegraphics[width=.23\textwidth]{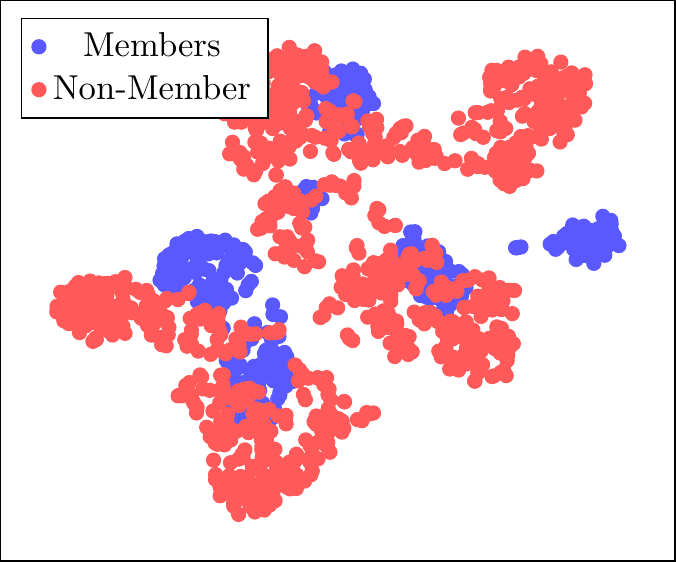}\hfill
    \includegraphics[width=.23\textwidth]{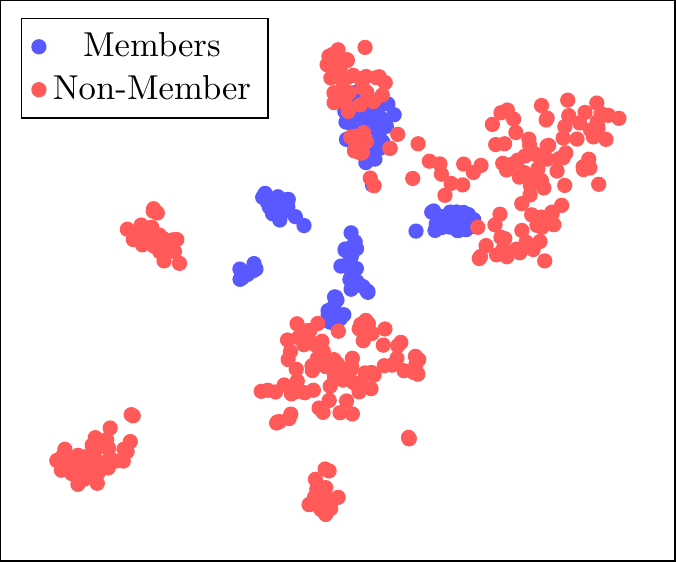}\hfill
\vspace{2mm}
    \includegraphics[width=.23\textwidth]{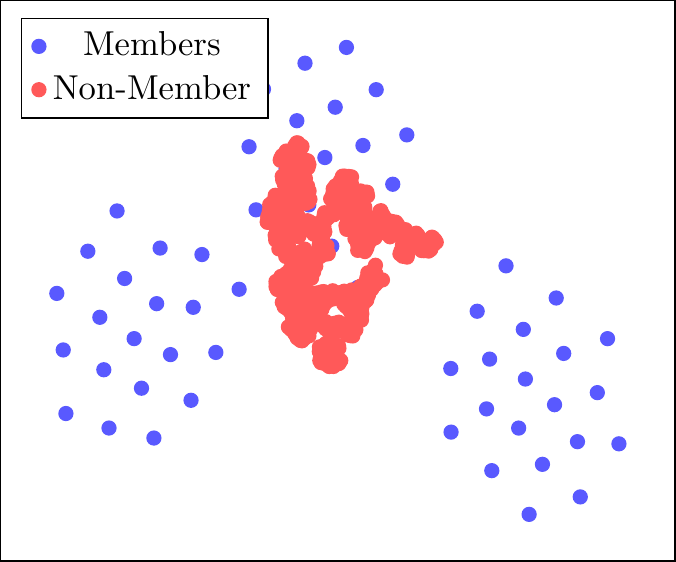}
\caption{Whitebox membership inference attacks exploit the distinguishable intermediate embedding of train and test graph nodes for Citeseer (left), Cora (right) and Pubmed dataset (bottom).}
\label{embedding}
\end{figure}

\noindent\textbf{Shadow Attack:} The shadow attack relies on the adversary training a local substitute model with similar functionality as the target model to identify characteristics of members and non-members.
The adversary has knowledge about the target GNN architecture and auxiliary graph dataset $G_{aux}$ sampled from the same underlying data distribution as $G_{train}$ which is consistent with prior attack settings~\cite{membershipinf,attributeinf,attributeinf2,logan} but the attack is transferable across different models~\cite{ndss19salem}.
This is a practical assumption where social networks have publicly available API enabling the adversary to obtain subgraphs of the original social network graph.
To conduct its attack, the adversary uses prior knowledge to map the target model's predictions to membership values and hence the attack is supervised.
For a target model $f()$, the adversary trains a substitute model $f_{local}$ on auxiliary graph data ($G_{aux}$) drawn from the same distribution as $G_{train}$.
The datasets are assumed to be non-overlapping, i.e, $G_{train} \cap G_{aux} = \phi$, which makes the attack more practical.
The goal is to train $f_{local}$ to mimic the behaviour of $f()$, i.e, the output predictions should be similar to each other $f_{local}(v;W') \sim f(v;W)$ for the same input user node $v$ but different parameters $W'$ and $W$ due to training on the different data.
Given the substitute model, the adversary creates a synthetic dataset with binary classes for distinguishing members and non-members (encoded as class 1 and class 0) of $f_{local}$'s training data $G_{aux}$ while using the output predictions as the input features.
That is, the synthetic dataset has the input as $f_{local}$'s predictions for an user node $v$ classified as "Member" if $v \in G_{aux}$ and "Non-Member" otherwise.
Hence, $f_{local}$ is used as a proxy for $f()$ to learn the mapping between the $f()$'s output predictions and the membership information.
The adversary trains a binary attack classifier $f_{attack}$ on the synthetic dataset used to predict whether a new user node was member of $G_{train}$.\\

\noindent\textbf{Confidence Attack:} In this particular case, we alleviate the assumption of adversary knowledge about data distribution and target model architecture as part of shadow model making the attack applicable to a wide range of practical scenarios.
Since, the adversary does not have prior knowledge to map the output predictions of target model to classify the membership, the attack is performed in an unsupervised setting. 
To conduct its attack, the adversary leverages the fact that graph nodes with higher output confidence prediction are likely to be members of $G_{train}$.
Here, the adversary finds the output prediction with highest confidence and compares whether this is above a certain threshold to decide whether the corresponding graph node was in the model's training graph $G_{train}$ or not.
A large output confidence indicates membership of the data point in the training data.
The adversary sweeps across different values to select the threshold value which best suits the application.
Prior work in traditional ML has indicated that the confidence attack is much better compared to shadow model attack as the which is verified in our experiments~\cite{princeton}.
The signal used to distinguish members from non-members from confidence score attacks is directly from the target model which the shadow model is subtle and uses a substitute model's output as a signal.
The attack success in this case is reliant on the quality of auxiliary data for training local substitute model and its functional similarity with the target model.


\subsubsection{Whitebox: Inference using Graph Embedding}

The adversary in a whitebox setting has access to the model's output for intermediate layers which in our case corresponds to the embedding for each node as an output of graph convolutional layer. 
This is a practical adversary assumption in the case of federated learning where intermediate computations can be observed~\cite{whitebox,collabinf}.

As explained Section~\ref{background}, GNNs compute the low dimensional embedding for the input graph data.
The parameters of GNNs are updated in each iteration of training and are tuned specifically for high performance on the train data resulting in a distinguishable footprint between embedding of train and test data points.
Figure~\ref{embedding} illustrates this rationale by plotting embedding of train and test graph nodes after a dimension reduction using 2D-TSNE algorithm~\cite{tsne}.

The attack methodology is unsupervised and the adversary has no supervised labels to map the intermediate embeddings to a membership value.
The adversary trains an encoder-decoder network in unsupervised fashion to map the intermediate embedding to a single membership value.
For an input graph node's embedding $\Psi (v)$, encoder $f_{enc}()$ generates a scalar membership value which is passed to decoder $f_{dec}(f_{enc}(\Psi (v)))$ to obtain $\Psi(v)$ by minimizing reconstruction loss: $||\Psi (v) - f_{dec}(f_{enc}(\Psi (v)))||_2^2$.
Given the membership values for different training and testing data points, K-Means clustering is used to cluster nodes into two classes (members and non-members).
In order to distinguish the clusters, the adversary requires to know the prior mapping of a small number of members and non-members.
For any new user nodes, the adversary then clusters them as members or non-members of the training data.

\subsection{Graph Reconstruction Attack}

Given the publicly released embeddings $\Psi (v)$ $\forall v \in G_{target}$  of sensitive target graph data ($G_{target}$), the goal of the adversary in this attack is to reconstruct $G_{target}$ and the corresponding connections between the different nodes $A_{target}$.
Specifically, the goal of the adversary is to reconstruct the adjacency matrix $A_{target}$ of the graph which is binary with $A_{ij}=1$ if there exists an edge between the node $i$ and $j$ and zero otherwise.
While node membership inference is a subtle privacy violation of user's data, this is a stronger attack where the entire sensitive graph data is reconstructed by the adversary.

Graph embeddings are specially computed to ensure that the underlying graph properties do not change.
In other words, the embeddings capture the rich semantic, invariant and structural information about the graph, for instance, by preserving proximity to the neighbouring nodes.
Hence, there exists a strong correlation between the released graph embeddings and the actual graph which can be exploited to reconstruct the graph data.

\begin{figure}[t]
    \centering
    \begin{minipage}[b]{1\linewidth}
    \centering

    \subfigure[Adversary trains attack model on auxiliary subgraph]{
   	\label{fig:mem_soft_label}
    \includegraphics[width=\linewidth]{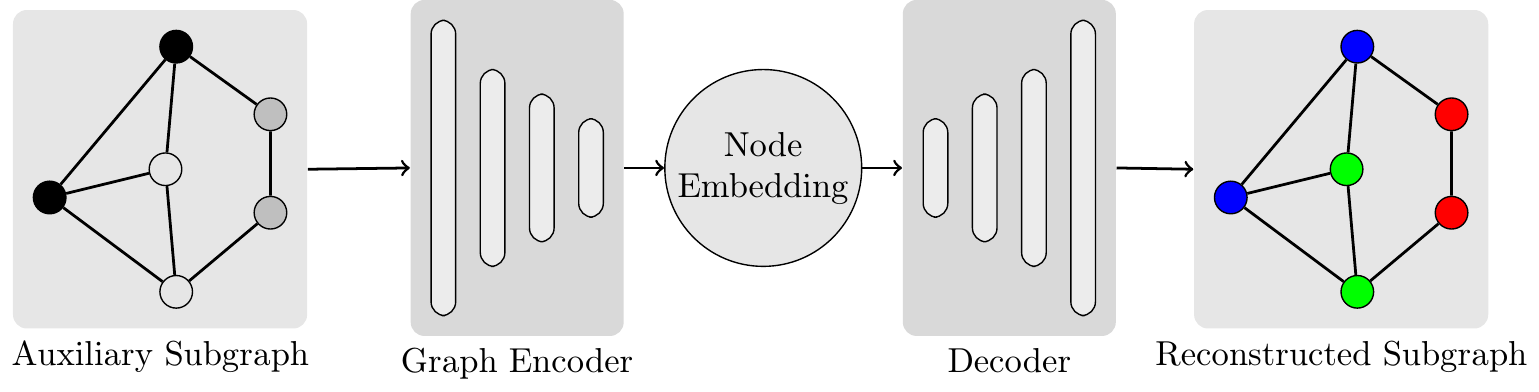}
    }

    \subfigure[Attack model reconstructs target graph]{
    \label{fig:mem_soft_label}
    \includegraphics[width=0.6\linewidth]{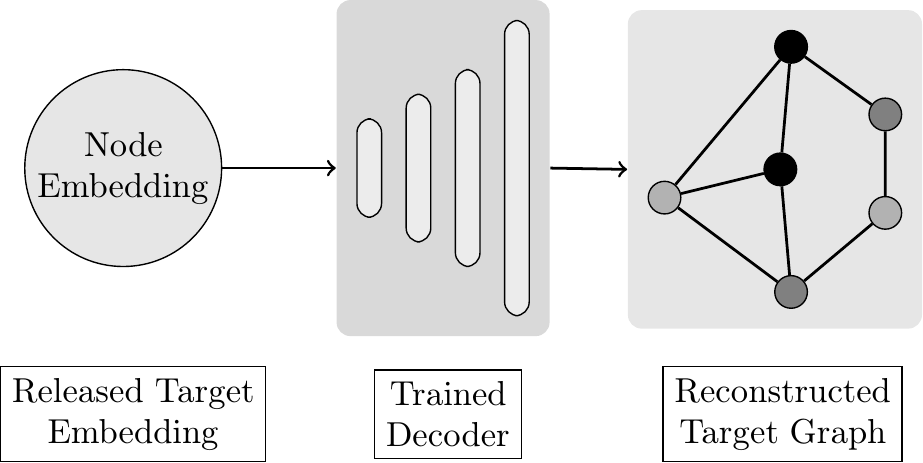}
    }
    \end{minipage}
    \caption{Attack methodology for graph reconstruction from released embeddings.}
    \label{fig:recattack}
\end{figure}

The adversary is assumed to have knowledge of the auxiliary subgraph $G_{aux}$ which is sampled from the same distribution as the target graph $G_{target}$.
Empirically, this is obtained by sampling two non-overlapping subgraphs from the full graph dataset.
The adversary performs graph reconstruction in two phases (Figure~\ref{fig:recattack}).
In Phase I, the adversary trains a graph encoder-decoder attack model on $G_{aux}$.
The graph encoder $f_{enc}$ maps the adjacency matrix of $G_{aux}$ to corresponding node embeddings $\Psi (v)\rightarrow f_{enc}(v)$ $\forall$ $v \in V$ represented as adjacency matrix.
The decoder $f_{dec}$ reconstructs the adjacency matrix $A_{rec} = f_{dec}(\Psi (v))$ while both the models are trained using backpropagation to minimize reconstruction loss: $||A - A_{rec}||_2^2$.
For the attack model, we consider an architecture with graph convolution as encoder and a decoder which computes the dot product between the embedding vector $\Psi (v)$ and its transpose $\Psi^T (v)$~\cite{Kipf2016tc}.
The attack models are trained on $G_{aux}$ and tested on the target embeddings corresponding to target graph $G_{target}$ to reconstruct the graph data.
Given the target released embeddings, the adversary then uses the trained decoder attack model to map the released embeddings to the target adjacency matrix $A_{target}^{rec} = f_{dec}(\Psi (v'))$ $\forall$ $v'\in G_{target}$.\\

\noindent\textbf{Link inference:}
Link inference is a direct result from the graph reconstruction attack where the adversary can check for an edge between two users using the adjacency matrix of the reconstructed graph.
This inference is a binary classification problem where the adversary aims to infer whether there exists a link between two nodes in the graph.
This inference represents the knowledge that two people know each other in online social networks for instance leading to identify the friendship circle of users which is a privacy violation of individuals.
More formally, given two user nodes $i$ and $j$, the adversary queries the reconstructed adjacency matrix $A_{target}^{rec}$ to infer whether there exists a link between $ij$ (if $A_{target}^{rec}[i][j] = 1$) or not (if $A_{target}^{rec}[i][j] = 0$).
The success of link inference attack closely depends on the success of reconstructing the target adjacency matrix $A_{target}^{rec} \sim A_{target}$.


\subsection{Attribute Inference Attack}

While the previous attack focussed on sensitive graph data and inference attacks exploiting connections between different nodes, attribute inference attacks exploit user node's sensitive features.
Particularly, given the embedding of the subgraph nodes and corresponding sensitive attribute ($\Psi (v), s_{v})$ $\forall v \in G_{aux}$, the adversary aims to infer the sensitive attributes $s^*$ corresponding to the publicly released target embeddings $\Psi (v')$ $\forall v' \in G_{target}$.
This is a practical assumption as a small fraction of users indeed make their information publicly available on their profile while other users prefer to keep such information private such as gender and location.

Nodes in graphs for most practical real world applications follow preferential connections, i.e, nodes similar to each other are connected to each other.
This is particularly true in case of social networks where users with similar likes and preferences, represented as features for nodes in the graph, are connected together~\cite{socialinfer,socialinfer2}.
This feature similarity and preferential connections in graphs are captured by graph embeddings to preserve the graph properties.
Hence, the embeddings are strongly correlated with the node features which can be exploited to infer sensitive attributes.

The adversary has access to the node embeddings and corresponding node's sensitive attributes ($\Psi (v), s_{v})$ $\forall v \in G_{aux}$ from the auxiliary subgraph known to the adversary.
The adversary uses this prior knowledge to train a supervised attack classifier $f_{attack}$ which maps the embedding to sensitive attributes, i.e, $f_{attack}: \Psi (v) \rightarrow s_{v}$.
Using this trained attack model, the adversary infers the sensitive attribute $s^*$ corresponding to the target embeddings $f_{attack}(\Psi(v'))$ where $v' \in G_{target}$.

\section{Experiment Setup}\label{setup}

In this section, we present the considered datasets, embedding algorithms, evaluation metrics and the methodology.

\subsection{Datasets}

For the membership inference and graph reconstruction attack, we consider three standard benchmarking datasets: Pubmed, Citeseer and Cora.
For the attribute inference attack, in turn, we consider two social networking datasets with anonymized sensitive attributes: Facebook\footnote{http://snap.stanford.edu/data/ego-Facebook.html} and LastFM\footnote{http://snap.stanford.edu/data/feather-lastfm-social.html}.

\noindent\textbf{Pubmed.} The Pubmed Diabetes dataset consists of 19,717 scientific publications from PubMed database pertaining to diabetes classified into one of three classes. The citation network consists of 44,338 links. Each publication in the dataset is described by a TF/IDF weighted word vector from a dictionary which consists of 500 unique words.
We use 60 train samples, 500 validation samples and 1,000 test samples.

\noindent\textbf{Citeseer.} The CiteSeer dataset consists of 3,312 scientific publications classified into one of six classes.
The citation network consists of 4,732 links.
Each publication is described by a 0/1-valued word vector indicating the absence/presence of the corresponding word from the dictionary.
The dictionary consists of 3,703 unique words.
The number of training samples is 120, 500 validation samples and 1,000 test samples.

\noindent\textbf{Cora.} The Cora dataset consists of 2,708 scientific publications classified into one of seven classes.
The citation network consists of 5,429 links. Each publication is described by a 0/1-valued word vector indicating the absence/presence of the corresponding word from the dictionary.
The dictionary consists of 1,433 unique words.
For training 140 samples are used, 300 validation samples and 1,000 test samples.

\noindent\textbf{Facebook.} The dataset comprises of 4,039 nodes representing different user accounts on the social network connected with each other through 88,234 edges.
Each user node has different features including the gender, education, hometown etc.
The user information has been anonymized through pseudonymization and the interpretation of the features have been obscured (i.e, attributes 'Male' and 'Female' have been replace with 'Gender 1' and 'Gender 2', respectively).

\noindent\textbf{LastFM.} The dataset was collected in March 2,020 using the public API provided by the social network specifically created for users from Asian countries.
The dataset has 7,624 nodes connected together with 27,806 edges based on mutual follower relationships.
Each user has attributes such as the music and artists they likes, location etc.

\subsection{Embedding Algorithms}

\noindent For the purpose of our experiments, we consider two classes of embedding algorithms: GNNs and random walk based.
We consider the following GNN based embedding techniques:

\noindent\textbf{Graph Convolutional Network (GCN)~\cite{Kipf2016tc}.} GCN computes the target node features from neighbouring nodes using matrix factorization, by normalizing adjacency matrix $A$ as $D^{-1}A$ where $D$ is the diagonal node degree matrix and results in averaging of neighbouring node features.
An additional trick is to use a symmetric normalization as $D^{-\frac{1}{2}}\hat{A}D^{-\frac{1}{2}}$.

\noindent\textbf{GraphSAGE~\cite{NIPS20176703}.} GraphSAGE extends the operations in GCN to more generic functions for transformation and aggregating of node features.
While the embedding of graph data in GCN relies on matrix factorization, GraphSAGE uses node feature aggregation using mean, LSTM and pooling to learn the embedding function.

\noindent\textbf{Graph Attention Networks (GAT)~\cite{velickovic2018graph}.} 
Weights associated with features during aggregation are explicitly defined and learnt during training.
GAT implicitly defines the weights using self-attention mechanism over the node features.

\noindent\textbf{Topology Adaptive GCN (TAGCN)~\cite{du2018topology}.} Instead of using the spectral convolutions for learning non-linear graph data, TAGCN proposes to use general K-localized filter convolution in the vertex domain.
It replaces the fixed square filters in traditional spectral GCNs for the grid-structured input data volumes.

\noindent For membership inference, we consider the embeddings from all the above four architectures for the whitebox setting while for the blackbox setting we consider only GraphSAGE algorithm as inductive training graphs models is challenging and GraphSAGE architecture is specifically designed to work in such training settings~\cite{NIPS20176703}.
In case of graph reconstruction attacks, we consider the generic GCN model as the encoder for the attack model.
In case of attribute inference attacks, we consider two state of the art unsupervised graph embedding algorithms based on random walk, namely, Node2Vec~\cite{node2vec} and DeepWalk~\cite{deepwalk}.

\noindent\textbf{DeepWalk~\cite{deepwalk}.} The algorithm creates a transition matrix from the graph and samples random walks from the matrix.
The nodes are viewed as words and the random walks are viewed as sentences and the resulting sequences are passed to Word2Vec and SkipGram~\cite{wordemb} to obtain node embeddings.

\noindent\textbf{Node2Vec~\cite{node2vec}.} This is an extension of DeepWalk which combines Breadth First and Depth First search explorations on the graph to create biased random walks.
The embeddings are computed using Word2Vec algorithm as mentioned above.

\subsection{Metrics}


\noindent To estimate the attack success of both membership and link inference, we consider the inference accuracy.

\noindent\textbf{Inference Accuracy.} Membership and link inference are binary classification problems: node is part of the training data or not (membership inference) and there exists a link between any two nodes or not (link inference).
Hence, the accuracy of random guess is 50\% and any higher accuracy indicates a privacy leakage about the model's sensitive training data.
In order to compute the additional benefit the adversary gets in terms of performing the attack over random guess, we name 'adversary advantage' a metric computed as: $I_{adv} = 2*(I_{acc}-0.5)$.
This metric estimates the information leakage from the model compared to a random guess.


\noindent For evaluating the performance of graph reconstruction attacks, we use two main metrics: precision and roc score.

\noindent\textbf{Precision.} The ratio of true positives is given by the precision and estimates the percentage of the predicted samples that are actually in the target graph.

\noindent\textbf{ROC-AUC Score.} The ROC curve plots the true positive rate on the y-axis and the false positive rate on the x-axis. The AUC score computes the area under the ROC curve to get how good the model distinguishes between different classes.
For a binary classification problem of graph reconstruction to obtain the binary adjacency matrix, the random guess accuracy is 50\% and any higher accuracy indicates the adversary's advantage in reconstructing the target graph. 

\noindent In case of attribute inference attack, we evaluate using the F1 score to balance both the recall and precision.

\noindent\textbf{F1-Score.} This metric computes the harmonic mean between the precision and recall which estimates the percentage of samples in the target graph which are predicted as such.

\subsection{Methodology}

In this work, we specifically focus on inductive training of GNN where the model does not see test nodes during training unlike transductive learning where the entire graph and features are available apriori.
Given the full graph $G_{full}$, we sample a subgraph $G_{train}$ which is used for training the models and evaluate the model performance on the held out graph $G_{full}-G_{train}$.
Such an inductive setting enables the adversary to learn new information about the target model's training graph resulting in a privacy leakage.

It is worth noting that none of the models have been overfitted on purpose. 
We consider the state of the art architectures and trained them using the state of the art libraries or by using training details as mentioned in their original papers. 
Consequently, the overfitting in the models is natural and not forced. 
In other words, the attacks are valid for all graph models including the state of the art models which are likely to be deployed for different services.

\section{Evaluation}\label{evaluation}

In this section, we evaluate the privacy leakage from attacks.

\subsection{Membership Inference Attack from Output Predictions}

\begin{figure}[!htb]
    \centering
    \begin{minipage}[b]{1\linewidth}
    \centering
    \subfigure[Generalization Error \hspace{0.5in} (b) Inference Advantage]{
    \label{fig:nonmem_soft_label}
    \includegraphics[width=0.5\linewidth,height=3.5cm, keepaspectratio]{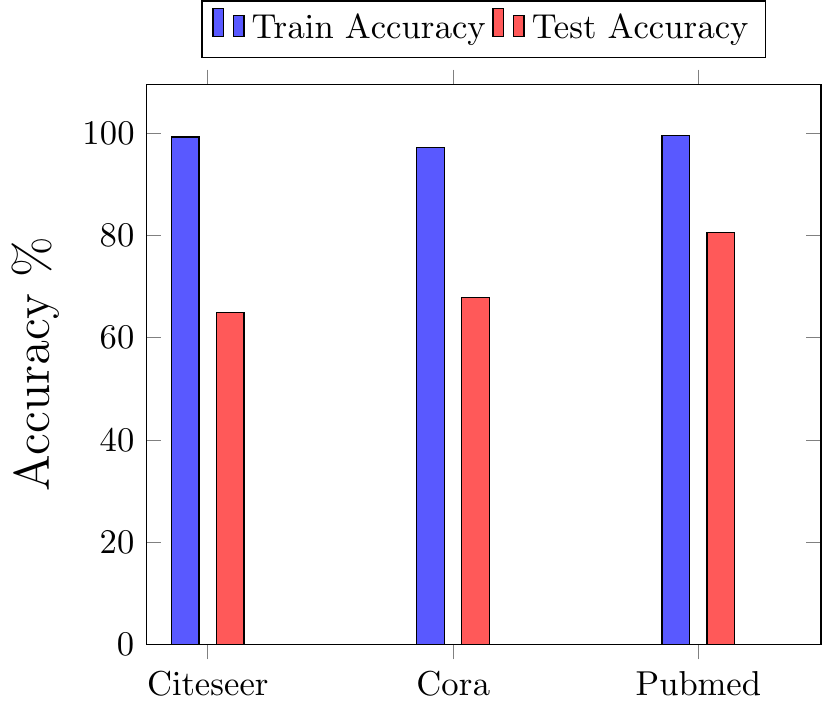}
    \includegraphics[width=0.5\linewidth,height=3.5cm, keepaspectratio]{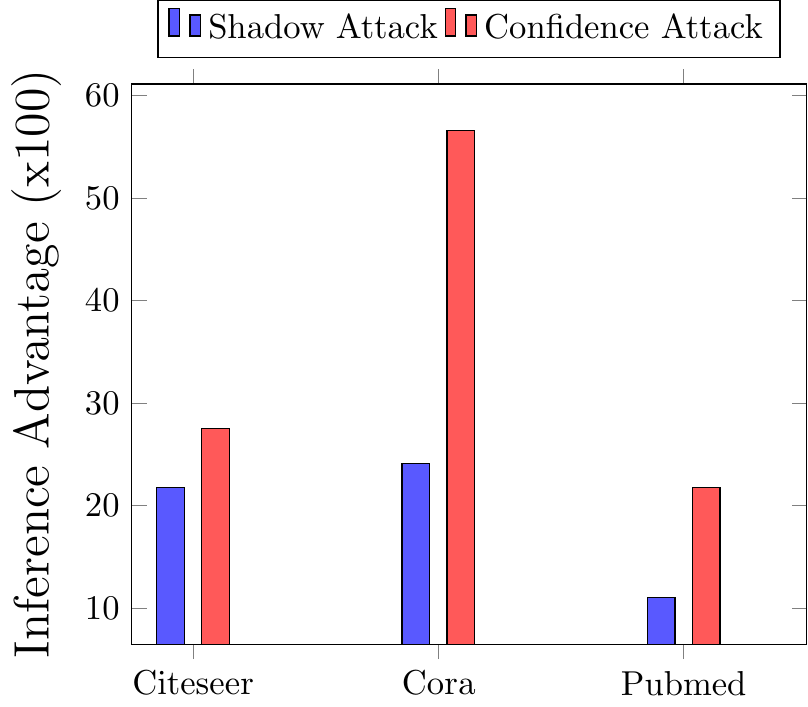}
    }
    \end{minipage}
    \caption{Blackbox membership inference attack uses the output predictions to give adversary an inference advantage.}
    \label{fig:NIA}
\end{figure}

The overfitting for GraphSage architecture trained on the three datasets is given in Figure~\ref{fig:NIA}(a).
We evaluate the two blackbox inference attacks exploiting the output predictions from the models: shadow inference which uses auxiliary knowledge on the data distribution and confidence inference which does not use auxiliary knowledge.
Results depicted Figure~\ref{fig:NIA}(b) show that under confidence inference attacks, the inference accuracy is 78.28\% (corresponding to an adversary's advantage of 27.48\%), 63.75\% (an adversary's advantage of 56.56\%), and 60.89\% (an advantage of 21.78\%) for Cora, Citeseer, and Pubmed datasets, respectively.
In case of shadow model attacks, the inference accuracy for Cora is 62.06\% (representing an adversary advantage of 21.74\%), 60.87\% for Citeseer (an adversary advantage of 24.12\%), 55.51\% for Pubmed dataset (an adversary advantage of 11.02\%).
Membership leakage is thus higher in confidence attack (i.e., without auxiliary knowledge) compared to shadow model attack (i.e., with auxiliary knowledge). While counter-intuitive, this result is consistent with similar attack methodology for traditional machine learning models~\cite{princeton}. The success of shadow model attack depends on the quality of the auxiliary dataset and the closeness of the shadow model to the target model. For graph models, we consider only one shadow model due to limited data size but the attack performance could improve at the cost of multiple shadow models trained on more data~\cite{membershipinf}.

\begin{figure}[!htb]
\centering
\includegraphics[width=0.4\textwidth,height=3.8cm]{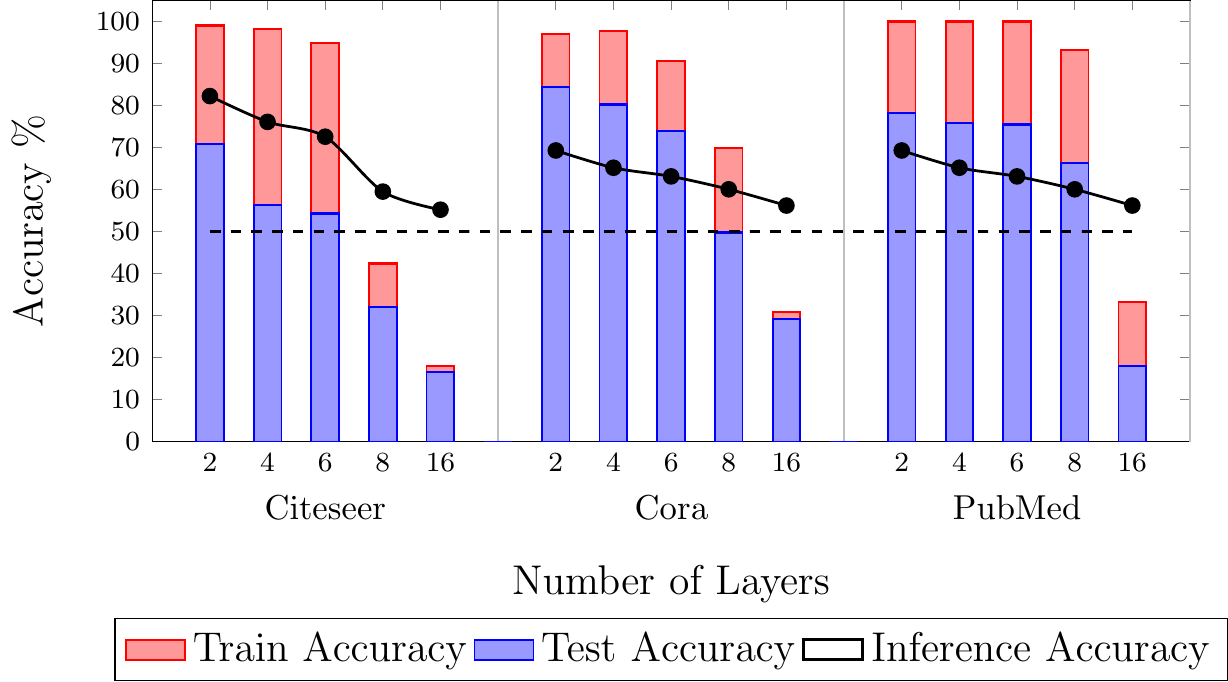}
\caption{The inference accuracy and predictive power decreases on increasing the number of layers due to feature over-smoothening from nodes deeper in the graph.}
\label{fig:numlayers}
\end{figure}

\noindent\textbf{Impact of Increasing Number of Layers.} We evaluate the performance of confidence attack on increasing the range of neighbourhood nodes used for aggregating the features (Figure~\ref{fig:numlayers}).
To do that, we extend the range of the message passing algorithm by increasing the number of layers in the GNNs~\cite{klicpera2018combining,Li2018DeeperII}.
On increasing the number of layers from 2 to 6, the inference accuracy decreases by 8\% for GCN. 
Interestingly, the generalization error increases (train accuracy remains the same but the test accuracy decreases) for Cora, Citeseer and Pubmed dataset, but the inference accuracy continues to decrease which indicates that the influence of preferential connections between different nodes in the graph plays a significant role in influencing the inference accuracy. 
For large number of layers ($>$ 8 layers) in the GNN, for all datasets and architectures, the model completely loses its predictive power.
In general, the inference accuracy as well as prediction accuracy decreases with increasing the range of the message passing algorithm by increasing the layers from 2 to 16.
This implies that the membership privacy leakage is influenced by the structured graph data with preferential connections between different nodes.
Specifically, aggregating features from larger number of nodes results in higher averaging which reduces the distinguishability (over-smoothening of features) as model converges to random walk’s limit distribution~\cite{klicpera2018combining,Li2018DeeperII} which is crucial for inference attacks~\cite{membershipinf,ndss19salem}.

\subsection{Membership Inference Attack from Graph Embeddings}

\begin{figure}[!htb]
  \begin{center}
    \includegraphics[width=0.5\linewidth]{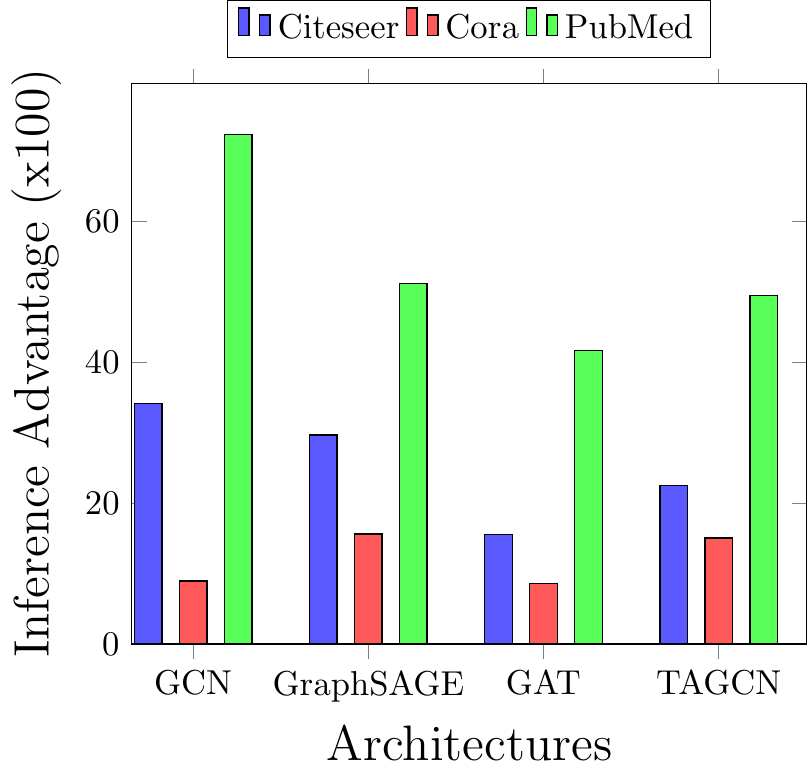}
  \end{center}
  \caption{Adversary advantage for node membership inference from Graph Embeddings.}
  \label{fig:whitebox}
\end{figure}

We exploit the difference in intermediate feature representation of train and test data to perform membership inference attack in a whitebox setting.
Results show that different models trained on PubMed dataset leak significantly more information between 20\% and 36\% over random guess accuracy.
On the other hand, models trained on Citeseer dataset provide to the adversary an advantage between ~7\% and ~17\% over random guess while for Cora dataset it is between 4\% and 7\%.
The embedding is significantly different for train and test data points for PubMed dataset as seen Figure~\ref{fig:whitebox} which result in a higher whitebox membership inference accuracy compared to Cora and Citeseer dataset.
The higher accuracy for Pubmed dataset can be attributed to significant distinguishability of features as seen by visually inspecting in Figure~\ref{embedding}.

\noindent The success of the unsupervised whitebox attack is attributed to the message passing which updates the parameters (weights) to specifically ensure higher distinguishability between the data points of different classes for high accuracy on training dataset.
Indeed, the parameters are specifically updated to fit the training dataset resulting in a high distinguishability between feature embedding of train and test data records.
Moreover, the feature embedding for the initial layers are useful since for later layers the features are oversmoothened which also reduces accuracy (as seen in increasing the range of nodes of message passing algorithm).

\subsection{Graph Reconstruction Attack}

\begin{figure}[!htb]
    \centering
    \begin{minipage}[b]{1\linewidth}
    \centering
    \subfigure[Validation ROC \hspace{1.2in} (b) Average Precision]{
   	\label{fig:mem_soft_label}
    \includegraphics[width=0.5\linewidth]{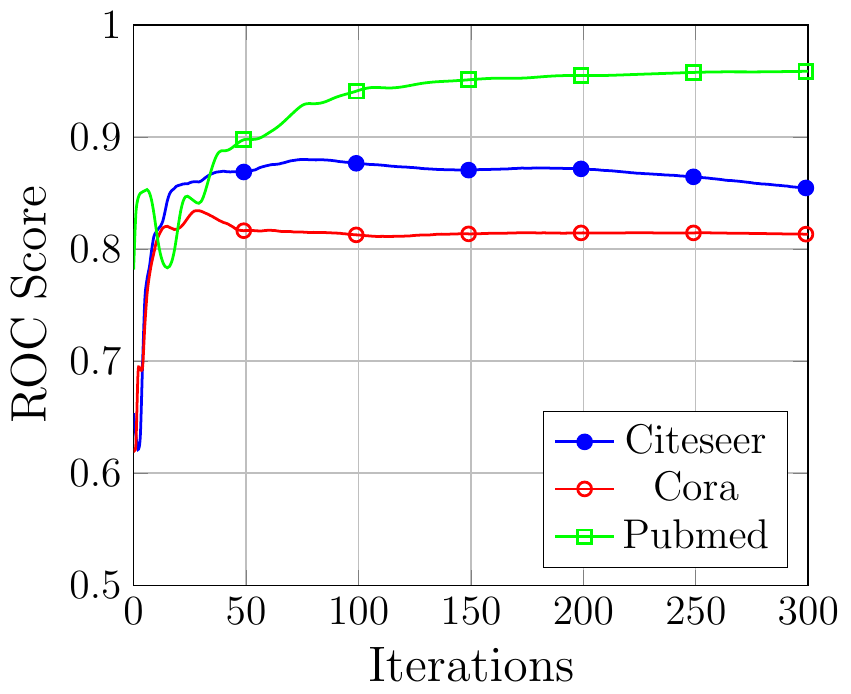}
    \includegraphics[width=0.5\linewidth]{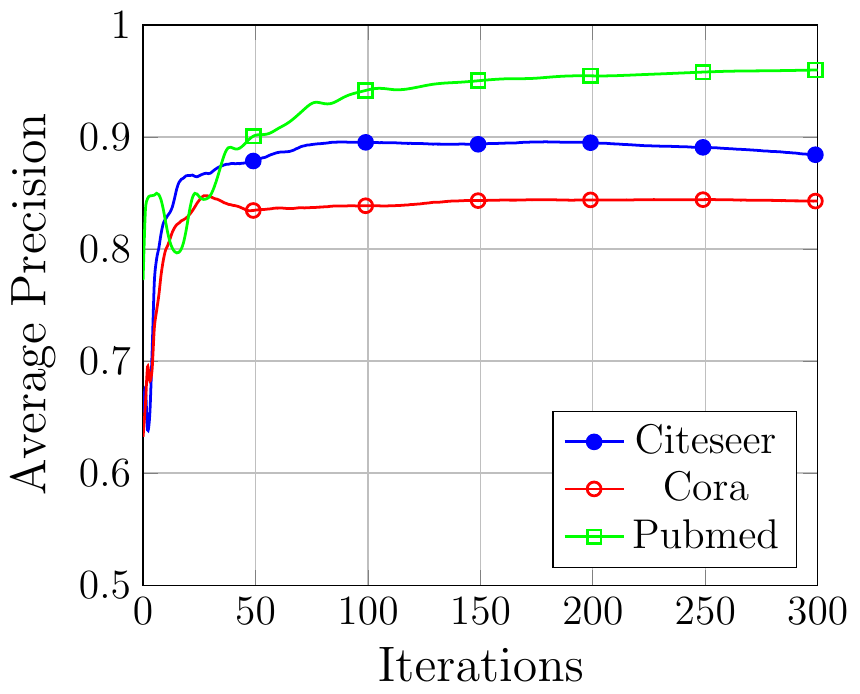}
    }
    \end{minipage}
    \caption{Training curves for AUC score and Average Precision on the validation graph.}
    \label{fig:valgraphrecon}
\end{figure}

The success of graph reconstruction is evaluated on the unseen target graph while the model is trained on the train graph (adversary's $G_{aux}$).
The test AUC score for Cora dataset is 0.65 and the average precision is 0.722 while for Citeseer dataset the AUC score is 0.65 and 0.778 average precision.
For Pubmed dataset, we get an average precision of 0.95 on the test set with an ROC of 0.94 in reconstructing the target test graph.
The curve for variation of AUC score and the average precision for the three datasets on the validation sub graph is given in Figure~\ref{fig:valgraphrecon}.

\noindent\textbf{Impact of Adversary Knowledge.} On increasing the adversary's knowledge to 50\% of the target graph, we observe an increase the attack performance.
Specifically, the AUC score for Cora increases to 0.76 from 0.65 while the average precision increases to 0.81 from 0.722.
In Citeseer dataset, the AUC score increases to 0.779 from 0.65 while the average precision increases to 0.828 from 0.778.
Finally, for Pubmed dataset showed an increase to 0.95 from 0.94 AUC score and 0.96 from 0.95 average precision.

\subsection{Link Inference Attack}

A direct implication of graph reconstruction attack is inferring whether there exists a link between two nodes in the network by querying the reconstructed adjacency matrix.
This is a binary classification problem.

\begin{figure}[!htb]
  \begin{center}
    \includegraphics[width=0.5\linewidth]{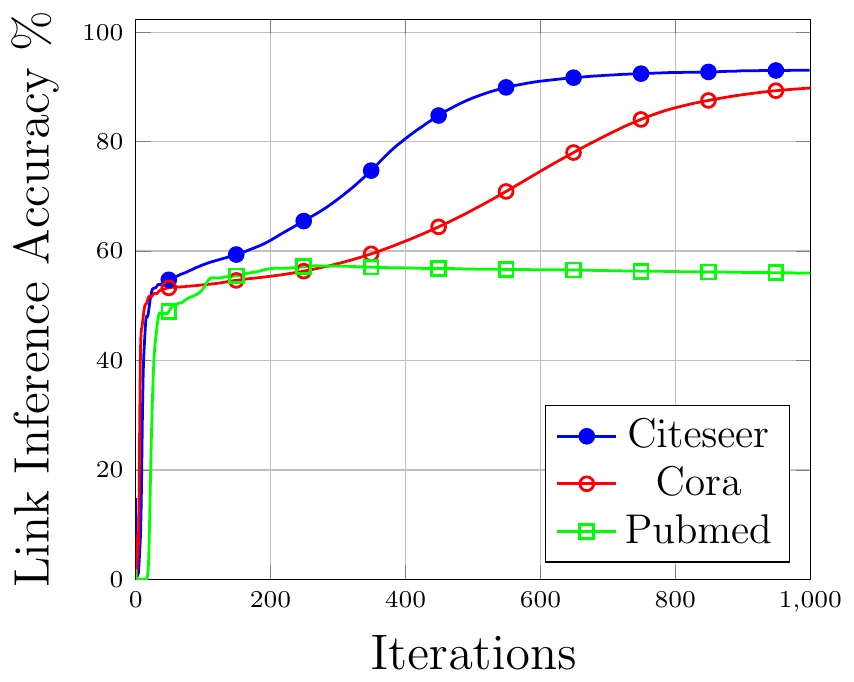}
  \end{center}
  \caption{Link Inference Accuracy Curve over different epochs}
\label{fig:lia}
\end{figure}

Figure~\ref{fig:lia} depicts the outcome of this attack for the three datasets.
For Citeseer dataset, the accuracy of of inference is around 93.39\% while for Cora dataset the inference accuracy is 90.73\% and 57.28\% for Pubmed dataset.
This indicates an adversary advantage of 86.78\% (Citeseer), 81.06\% (Cora) and 14.56\% (Pubmed).
The same train-test-validation distribution is used for the three datasets with 30\% train records and 60\% test records and remaining 10\% as validation records.

\subsection{Attribute Inference Attack}

In case of attribute inference attacks, we evaluate two state of the art embedding models: Node2Vec and DeepWalk, using three attack models: Neural Networks (NN), Random Forest (RF) and Support Vector Machine (SVM).
We generate embeddings using the two algorithms on Facebook and LastFM dataset which contain gender and location as sensitive attributes, respectively.
That is, the adversary infers user gender as a target sensitive attribute in Facebook dataset classified into one of three classes: Male, Female and Others.
The location target attribute for LastFM dataset is categorized in 18 locations for the users in the network.

\begin{figure}[!htb]
    \centering
    \begin{minipage}[b]{1\linewidth}
    \centering
    \subfigure[LastFM \hspace{1.2in} (b) Facebook]{
   	\label{fig:mem_soft_label}
    \includegraphics[width=0.5\linewidth,height=3.5cm, keepaspectratio]{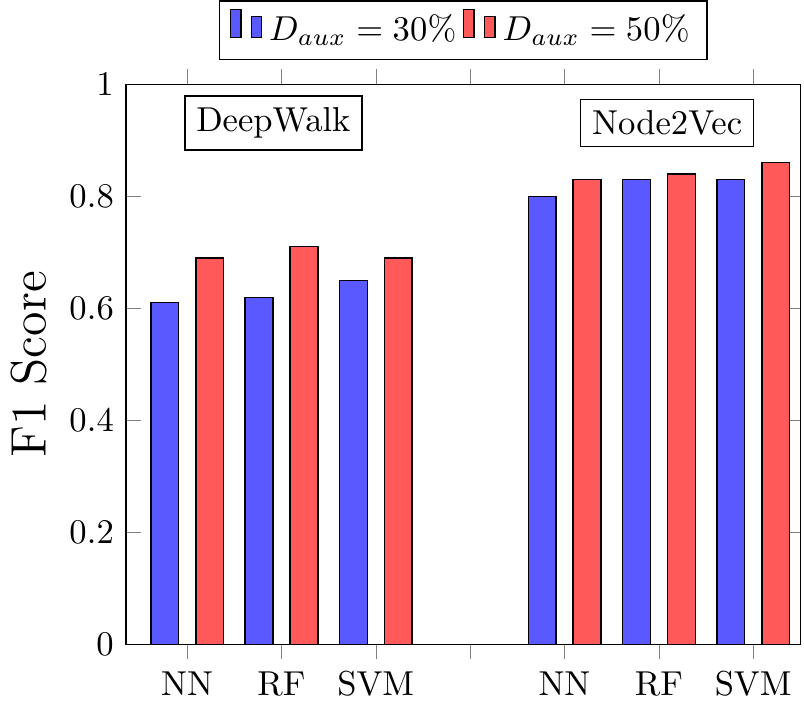}
    \includegraphics[width=0.5\linewidth,height=3.5cm, keepaspectratio]{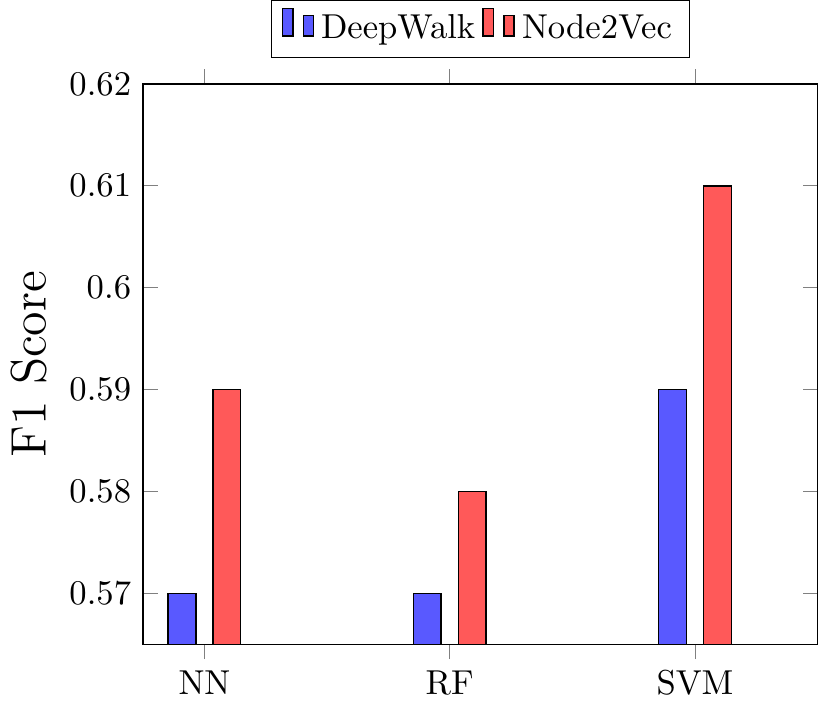}
    }

    \end{minipage}
    \caption{F1 score for different attack classifiers to infer sensitive attributes.}
    \label{fig:aia}
\end{figure}

The inference attack performance is given by the F1 score (Figure~\ref{fig:aia}). 
For Facebook, the graph embedding using DeepWalk resulted in an F1 score of 0.57 for NN, 0.58 for RF and 0.59 for SVM classifier.
On the other hand, Facebook's Node2Vec embedding showed an F1 score of 0.59, 0.57 and 0.61 respectively for NN, Rf and SVM attack classifier.
In case of LastFM, we found the attack F1 scores for Node2Vec for higher than DeepWalk embeddings.
The F1 score for DeepWalk 0.61, 0.62 and 0.65 corresponding to NN, RF and SVM attack classifier while the F1 score using Node2Vec embeddings are 0.80, 0.83 and 0.83 for NN, RF and SVM.

\noindent\textbf{Impact of Adversary's Knowledge.} The performance of the attack model for Facebook dataset did not increase by much.
On increasing the knowledge of the adversary's auxiliary dataset from 30\% to 50\%, the confidence of attack on LastFM dataset increases.
For DeepWalk algorithm, the attack F1 score increases to 0.69 from 0.61 for NN, 0.71 from 0.62 for RF and 0.69 from 0.65 for SVM attack classifier.
On the other hand, for Node2Vec, the attack model F1 score increased to 0.83 from 0.80 for NN, 0.84 from 0.83 for RF and 0.86 from 0.83 for SVM.

\section{Related Work}
\label{related}

The wide availability of location and mobility data has been followed by the development of machine learning algorithms to predict interests, colocations or other sensitive user information~\cite{noulas2009inferring}. 
Traditional approaches define features to characterize users’ data and use the features for various prediction tasks. But, an important pre-processing step for using graph data with machine learning is embedding the high dimensional graph data to a low dimensional representation for easy processing by machine learning algorithms~\cite{yang2019revisiting}.
In this context, GNNs~\cite{zhou2018graph} have state of the art performances on complex graph data for tasks including node classification and link prediction.
However, the privacy implications of the use of such embeddings have not been fully considered.

Inference attacks that violate data privacy have been explored in the context of traditional machine learning models.
Membership Inference attacks can be deployed in both whitebox~\cite{whitebox} and blackbox~\cite{membershipinf} setting in traditional machine learning algorithms.
These attacks are further extended to collaborative learning~\cite{collabinf,whitebox} and generative models~\cite{logan}.
On the other hand, reconstruction attacks infer private attributes of the inputs passed to the models~\cite{attributeinf, attributeinf2, propertyinf, modelinversion}.
Other privacy attacks aim to extract hyperparameters~\cite{8418595}, reverse engineer the model architecture and parameters using side channels~\cite{timing} or the output predictions~\cite{stealml}.
Memorization of data by Neural Networks has been attributed as a major cause for privacy leakage~\cite{memorize,secretsharer,overlearninginf}.
Further, recent works have indicated privacy risks in Graph NNs where an adversary can infer the presence of a link between two nodes using a manual threshold between the distance of two node features~\cite{linksteal}.
This attack however, is subsumed within our more generic attack methodology where we extract the entire adjacency matrix which can be used to infer the presence of links.
Text models have been shown to leak user data through attribute inference and inversion attacks~\cite{textembleak,nlp}. 
However, a direct application of these attacks is not possible in case of high dimensional graphs and requires additional consideration to the network structure making our problem challenging.
Other than privacy attacks, adversarial attacks against GNNs~\cite{graphatt,nodepoison} have been explored as well as training algorithms to enhance the robustness against such attacks~\cite{robustdef1,robustdef2}.

To mitigate membership and attribute inference attacks, Memguard~\cite{memguard} and AttriGuard~\cite{attriguard} add carefully crafted noise to the final output prediction to misclassify the shadow model attacks.
Distillation for membership privacy~\cite{shejwalkar2021membership} is the state of the art defense  that uses knowledge transfer to regularize its models by carefully generating transfer set used for knowledge transfer. 
Adversarial regularization using minimax optimization regularizes the model to mitigate inference attacks~\cite{advreg}.
\cite{ndss19salem} study regularization based defenses that use ensemble training, dropout, and L2-regularization.
Differential Privacy mitigates such privacy attacks with theoretical guarantees by adding noise to gradients but faces an unbalanced privacy accuracy trade-off~\cite{diffpriv}.
Such Differential Privacy frameworks have also been explored in the context of graph and text embeddings~\cite{dptext,dpne} but their efficacy on lowering privacy risks from the proposed attacks is yet to be explored.

\vspace{-2mm}
\section{Discussions and Conclusions}
\label{conclusions}

This work provides the first comprehensive privacy risk analysis related to graph embedding algorithms trained on sensitive graph data.
Specifically, this paper quantifies privacy leakage of three major classes of privacy attacks under practical adversary assumptions and threat models, namely membership inference, graph reconstruction and attribute inference.
Firstly, an adversary conducting a membership inference attack aims to infer whether a given user's node was used in the training graph dataset or not.
Secondly, publicly released embeddings can be inverted to obtain the input graph data enabling an adversary to perform graph reconstruction attack on the sensitive graph data.
This further enables the adversary to perform link inference attack to infer whether a link exists between two nodes in the network.
Finally, we show that an adversary can infer sensitive hidden attributes of users such as gender and location from the graph embeddings.
Our results underlines many privacy risks in graph embeddings and calls for further research to mitigate these privacy threats.

Potential mitigation strategies to lower the privacy risks can be considered.
For instance, lowering the precision of the embedding vector for each node by rounding can help to reduce the attack model from learning rich features about the inputs~\cite{membershipinf,nlp}.
In the proposed attacks, the attacker model is a machine learning algorithm vulnerable to adversarial examples, i.e, imperceptible noise added to the output prediction to force the target model to misclassify.
The embeddings can be released with an additional adversarial noise to misclassify the target model while additionally ensuring utility~\cite{attriguard,memguard}.
Further, the inference attacks can be modelled within the training process as a minimax adversarial training with joint optimization to minimize the model loss using the graph embeddings (e.g., GNNs) while maximising the adversary's loss on inferring the sensitive inputs~\cite{advreg,textembleak}.
Finally, Differential Privacy can provide a theoretical bound on the total privacy leakage from the downstream processing from embeddings on an individual's data point~\cite{dptext,dpne}.
However, the efficacy of these potential mitigations are left for future work.

{\footnotesize
\bibliographystyle{ACM-Reference-Format}
\bibliography{paper.bib}
}


\end{document}